\documentclass[sigconf]{acmart}

\usepackage{booktabs} 
\usepackage{soul}

\usepackage[detect-all]{siunitx}
\usepackage{amsmath}
\usepackage{amssymb}
\usepackage[ruled,vlined]{algorithm2e}
\usepackage{booktabs} 
\usepackage{multirow}
\usepackage{enumitem}

\usepackage{amsmath,amssymb,amsfonts}
\usepackage{algorithmic}
\usepackage{graphicx}
\usepackage{textcomp}
\usepackage{xcolor}

\usepackage[nodisplayskipstretch]{setspace}

\SetAlFnt{\small}

\usepackage{hyperref}
\hypersetup{draft}

\usepackage[framemethod=tikz]{mdframed}

\usepackage{esvect}

\usepackage[utf8]{inputenc}
\usepackage{listings}
\usepackage{xcolor}
\definecolor{codegreen}{rgb}{0,0.6,0}
\definecolor{codegray}{rgb}{0.5,0.5,0.5}
\definecolor{codepurple}{rgb}{0.58,0,0.82}
\definecolor{backcolour}{rgb}{0.95,0.95,0.92}

\lstdefinestyle{mystyle}{
    backgroundcolor=\color{backcolour},   
    frame=single,     
    commentstyle=\color{codegreen},
    keywordstyle=\color{magenta},
    numberstyle=\tiny\color{codegray},
    stringstyle=\color{codepurple},
    basicstyle=\ttfamily\scriptsize,
    breakatwhitespace=false,         
    breaklines=true,                 
    captionpos=b,                    
    keepspaces=true,                 
    numbers=left,                    
    numbersep=5pt,                  
    showspaces=false,                
    showstringspaces=false,
    showtabs=false,                  
    tabsize=2
}
\DeclareSIUnit\bit{b}
\DeclareSIUnit\byte{B}
\DeclareSIUnit\cycle{cycle}
\DeclareSIUnit\pixel{pix}

\DeclareMathOperator*{\minimize}{minimize}
\DeclareMathOperator*{\proj}{proj}

\newcommand{\embf}[1]{\emph{\textbf{#1}}}

\newcommand{\reffig}[1]{Fig.~\ref{#1}}
\newcommand{\refsec}[1]{\S\ref{#1}}
\newcommand{\reftab}[1]{Table~\ref{#1}}

\newcommand{\refeqn}[1]{Eqn.~\ref{#1}}

\newcommand{\hw}[1]{\ensuremath{\mathtt{#1}}}
\newcommand{\hws}[2]{\ensuremath{\mathtt{#1}_\mathrm{#2}}}
\newcommand{\csbengine}{\hw{CSB}-\hw{Engine} }

\newcommand{\ra}[1]{\renewcommand{\arraystretch}{#1}} 

\setcopyright{acmcopyright}
\definecolor{xiaolong}{RGB}{181, 97, 2}

\usepackage{hyperref}
\hypersetup{draft}

\copyrightyear{2020}
\acmYear{2020}
\setcopyright{usgovmixed}\acmConference[ICS '20]{2020 International Conference on Supercomputing}{June 29-July 2, 2020}{Barcelona, Spain}
\acmBooktitle{2020 International Conference on Supercomputing (ICS '20), June 29-July 2, 2020, Barcelona, Spain}
\acmPrice{15.00}
\acmDOI{10.1145/3392717.3392749}
\acmISBN{978-1-4503-7983-0/20/06}

\begin{document}

\title{CSB-RNN: A Faster-than-Realtime RNN Acceleration Framework with Compressed Structured Blocks}

\author[R. Shi, P. Dong, T. Geng, Y. Ding, X. Ma, H.K.-H. So, M. Herbordt, A. Li]{Runbin Shi$^{1,2,^*,\dagger}$, Peiyan Dong$^{2,\dagger}$, Tong Geng$^{3,\dagger}$, Yuhao Ding$^1$, Xiaolong Ma$^2$,}
\author[Y. Wang]{Hayden K.-H. So$^1$, Martin Herbordt$^3$, Ang Li$^4$, Yanzhi Wang$^2$}
\thanks{$^*$Contribution during visit at Northeastern University.\\$^\dagger$Equal contribution.}
\affiliation{%
  \institution{$^1$The University of Hong Kong, $^2$Northeastern University, $^3$Boston University, $^4$Pacific Northwest National Laboratory}
}
\email{{rbshi,yhding,hso}@eee.hku.hk, {dong.pe,ma.xiaol}@husky.neu.edu, {tgeng,herbordt}@bu.edu, ang.li@pnnl.gov} 
\email{yanz.wang@northeastern.edu}

\renewcommand{\authors}{Runbin Shi, Peiyan Dong, Tong Geng, Yuhao Ding, Xiaolong Ma, Hayden K.-H. So, Martin Herbordt, Ang Li and Yanzhi Wang}

\begin{abstract}
Recurrent neural networks (RNNs) have been widely adopted in temporal sequence analysis, where realtime performance is often in demand. 
However, RNNs suffer from heavy computational workload as the model often comes with large weight matrices. 
Pruning (a model compression method) schemes have been proposed for RNNs to eliminate the redundant (close-to-zero) weight values.
On one hand, the non-structured pruning methods achieve a high pruning rate but introducing computation irregularity (random sparsity), which is unfriendly to parallel hardware.
On the other hand, hardware-oriented structured pruning suffers from low pruning rate due to restricted constraints on allowable pruning structure. 

This paper presents CSB-RNN, an optimized full-stack RNN framework with a novel compressed structured block (CSB) pruning technique. 
The CSB pruned RNN model comes with both fine pruning granularity that facilitates a high pruning rate and regular structure that benefits the hardware parallelism.
To address the challenges in parallelizing the CSB pruned model inference with fine-grained structural sparsity, 
we propose a novel hardware architecture with a dedicated compiler.
Gaining from the architecture-compilation co-design, the hardware not only supports various RNN cell types, but is also able to address the challenging workload imbalance issue and therefore significantly improves the hardware efficiency (utilization). 
Compared to the vanilla design without optimizations, the hardware utilization has been enhanced by over $2\times$. 
With experiments on $10$ RNN models from multiple application domains, CSB pruning demonstrates $3.5\times$-$25\times$ lossless pruning rate, which is $1.6\times$ to $3.9\times$ over existing designs.
With several other innovations applied, the CSB-RNN inference can achieve faster-than-realtime latency of $0.79\mu$s-$6.58\mu$s in an FPGA implementation, which contributes to $1.12\times$-$12.57\times$ lower latency and $3.53\times$-$58.89\times$ improvement on power-efficiency over the state-of-the-art.
\end{abstract}

\ccsdesc[500]{Computer systems organization~Parallel architectures}

\keywords{RNN, structured pruning, workload balancing, FPGA}

\maketitle

\section{Introduction}
RNNs have been widely adopted for its high-accuracy on temporal sequence analysis, such as machine translation~\cite{cho2014learning}, speech recognition~\cite{graves2013speech}, or even stock-price prediction~\cite{selvin2017stock}.
However, the increasingly large model size and tremendous computational workload of the RNN hampers its deployment on embedded (edge) devices, which strictly demand realtime processing with limited hardware resources.
To address this issue, weight pruning techniques~\cite{lym2019prunetrain, han2015deep, wen2018learning} have been proposed, which shrink the model size, reduce storage demand, and provide higher potential hardware performance by eliminating the redundant (close-to-zero) weights and the corresponding arithmetic operations in inference.

The weight pruning schemes in some existing works \cite{han2015deep, shi2019lstm} are in a \emph{non-structured} fashion and with \emph{extreme irregularities} in the computation, which is unfriendly to either the modern parallel device or the hardware architecture design. Thus the performance degradation from the hardware inefficiency encroaches upon the gains from model compression. 
Therefore, researchers start to explore other pruning approaches, i.e., \emph{structured} pruning \cite{wen2018learning, mao2017exploring, narang2017block}, in which the regular computation patterns are maintained. 
Although these structured-pruned models are relatively hardware-friendly, the coarse pruning granularity (structure) leads to either a significant degradation on model accuracy or a limited pruning rate (the weight count ratio of the original model to pruned model). 
To keep the accuracy loss acceptable, the attainable pruning rates delivered in the existing structured pruning schemes are \emph{far lower than} that the ones with the non-structured pruning, wasting potential pruning opportunities.

In this paper, we aim to overcome the above limitations. We first propose a novel \emph{fine-grained structured} pruning technique (CSB pruning) that provides a similar compression rate (and test accuracy) as non-structured pruning while offering a higher potential for hardware acceleration than the non-structured methods. 
During the training phase, each weight matrices are divided into \emph{fine-grained blocks}, and a structured pruning is conducted on every block independently. The pruned blocks are encoded in a novel \emph{compressed structured block (CSB)} sparse format for inference acceleration, which significantly reduces the weight storage demand while retaining the fine-grained content in the original model. 

To realize a realtime inference with parallel hardware, there are still multiple challenges to design an architecture that can exploit the benefits of CSB pruning in a seamless manner.
In particular, the parallel architecture should handle massive fine-grained blocks with imbalanced workloads (sparsity) but maintain a high hardware efficiency (utilization).
Meanwhile, the architecture should be programmable for various RNN cell types (e.g., LSTM~\cite{hochreiter1997long}, GRU~\cite{cho2014learning}), although the existing RNN architectures are designed for a particular cell type. 
To address the issues above, we propose an architecture-compilation co-design to realize the best flexibility and acceleration performance. 
A programmable RNN dataflow architecture is designed that supports existing RNN cell types. 
In particular, the \csbengine in our architecture is designed with a novel \emph{workload sharing} technique.
With the one-shot compilation, the workload is automatically balanced among processing elements (PEs) in \hw{CSB}-\hw{Engine}, which improves the hardware efficiency to a near theoretical value.

\begin{figure}
  \centering
  \includegraphics[width=0.5\textwidth]{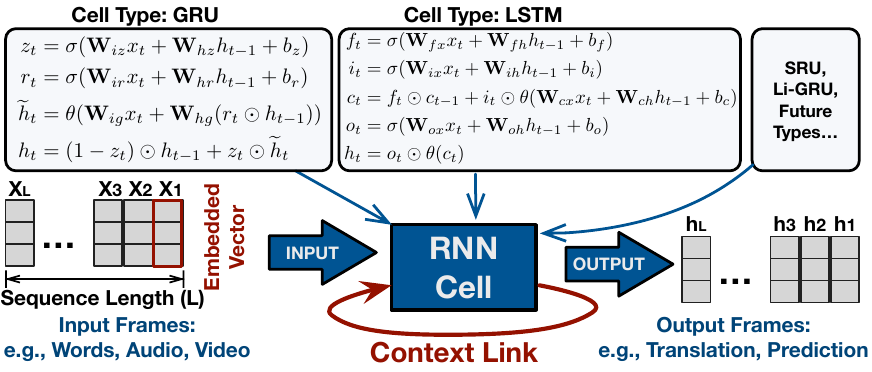}
  \caption{Computation flow of RNN inference. Note that there are multiple RNN cell types. The main workload is matrix-vector multiplication (MVM).}
  \label{fig:02_rnn}
\end{figure}

The major \embf{contributions} are summarized as follows:
\begin{itemize}[leftmargin=*]
  
  \item We present CSB-RNN, an optimized full-stack RNN acceleration framework, which facilitates running various types of RNNs with faster-than-realtime latency. CSB-RNN includes three innovations: (1) an adaptive and fine-grained structured compression technique, CSB pruning; (2) a programmable RNN dataflow architecture equipped with \hw{CSB}-\hw{Engine}; (3) a compiler design with optimizations to achieve almost perfect workload balance. 
  
  \item The proposed CSB pruning technique provides ultra-high ($3.5\times$-$25\times$) pruning rates without any loss on accuracy. Furthermore, CSB pruning does not incur high-degree computational irregularities, making highly efficient hardware acceleration possible.
  \item An architecture-compilation co-design is proposed to sufficiently exploit the benefits of CSB pruning and provide close-to-theoretical peak performance with automatic workload balancing.
  
  \item With experiments on $10$ RNN models from various application domains, CSB pruning demonstrates $3.5\times$-$25\times$ lossless pruning rate, which is $1.6\times$ to $3.9\times$ over existing designs. With the proposed architecture-compilation co-design applied, the CSB-RNN delivers faster-than-realtime inference with the latency of $0.79\mu$s-$6.58\mu$s in an FPGA implementation. The proposed framework contributes to $1.12\times$-$12.57\times$ lower latency (with even fewer computation resources) and $3.53\times$-$58.89\times$ improvement on power-efficiency over the state-of-the-art.
  
\end{itemize}

\section{Background}
\subsection{Temporal Sequence Processing with RNN}

The recurrent neural networks (RNNs) deliver high accuracy in the temporal sequence processing.
A typical schematic of RNN computation is depicted in \reffig{fig:02_rnn}.
Successive frames (e.g., word, phoneme) from the temporal sequence (e.g., sentence, voice) are \emph{embedded} as input neuron-vectors ($x_t$), and then sent to RNN cells for inference computation. 
$t$ represents the time point.
The output neuron-vector ($h_t$) contains the inference results (e.g., translation, prediction) that may have different dimensions with $x_t$.

Multiple RNN cell types exist that are composed of different \emph{computational dataflow} but almost the same \emph{arithmetic primitives}.
\reffig{fig:02_rnn} lists the arithmetic of two widely-used RNN cells, GRU~\cite{cho2014learning} and LSTM~\cite{hochreiter1997long}.
The significant workload is \emph{matrix-vector multiplication} (MVM) between the weight matrices and input/hidden neurons; And the rest workload is element-wise operations, including Sigmoid ($\sigma$)/Tanh ($\theta$) activation function, element-wise multiplication ($\odot$) and addition.
In particular, the RNN cell computation at time $t$ invokes the intermediate vector $c_{t-1}$ and output vector $h_{t-1}$ from the previous timestamp.
The data dependency results in a \emph{context link} between two successive RNN cell iterations, which helps to memorize the temporal feature during inference.

\subsection{RNN Weight Pruning Techniques}
\subsubsection{Non-structured Pruning v.s. Structured Pruning}\label{sec:diff_prune_scheme}

\begin{figure}
  \centering
  \includegraphics[width=0.42\textwidth]{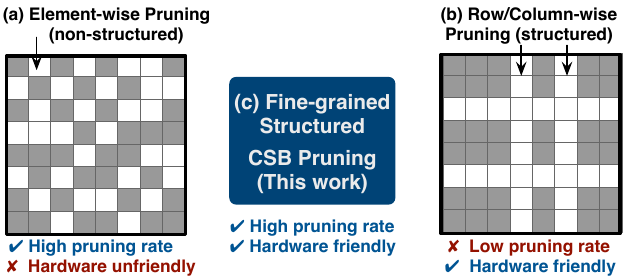}
  \caption{CSB pruning takes advantage of both non-structured (random) pruning (a) and coarse-grained structured (row/column) pruning (b).}
  \label{fig:02_prune}
\end{figure}

The pruning technique has been proposed for deep learning models to reduce redundant (close-to-zero) weights and thus the computation workload. 
The early non-structured pruning~\cite{han2015deep} achieves a high pruning rate; however, the random sparse model (\reffig{fig:02_prune}(a)) brings a high degree of irregularity to the inference computation, which is unfriendly to either the modern parallel device or the hardware architecture design.
Some existing works~\cite{han2017ese, cao2019efficient} address this issue by pruning model with \emph{region-balanced sparsity} (between non-structured and structured sparsity), which reduced the attainable pruning rate.
As \reffig{fig:02_prune}(b), the structured pruning schemes~\cite{wen2016learning, gao2018deltarnn} were proposed for hardware friendly purpose that the entire row/column is removed as a whole in pruning.
Although the pruned model maintains the regularity and can even be compacted to a dense matrix, the pruning rate with this scheme is relatively low due to the coarse pruning granularity.
With the advantages of both the non-structured and coarse-grained structured pruning methods, the CSB pruning in this work is a fine-grained structured method that not only achieves a high pruning rate but also makes the hardware acceleration possible.

\subsubsection{Model Training with ADMM-based Pruning Technique}\label{sec:admm_prune}
The training process is to find a proper set of weight values that reach the minimal classification loss compared to the ground truth. 
The objective of training an $N$-layer RNN can be formulated as,
{\small
\begin{equation}
\begin{aligned}
  &\minimize_{\{ \mathbf{W}_i\}, \{\mathbf{b}_i\}} &f(\{ \mathbf{W}_i\}_{i=1}^N, \{\mathbf{b}_i\}_{i=1}^N) \\
  &s.t. &\mathbf{W}_i \in \mathbf{S}_i, i=1,...,N
\end{aligned}
\label{eqn:general_train}
\end{equation}
}
where function $f$ represents inference loss on the given dataset, $\mathbf{S}_i$ is the feasible set of $\mathbf{W}_i$, which is subject to the user constraints. 
In the regular RNN training, $\mathbf{S}_i$ is $\mathbb{R}$ (i.e., no constraint), and thus the optimal weights ($\mathbf{W}_i$) and bias ($\mathbf{b}_i$) for each layer can be obtained by classical stochastic gradient descent (SGD) method~\cite{bottou2010large}.
However, once the \embf{weight pruning} is conducted along with the training process, the constraint of weight-sparsity represented by $\mathbf{S}_i$ becomes combinatorial and no longer convex, which prevents the \refeqn{eqn:general_train} from being solved by classical SGD. 
The advanced \emph{Alternating Direction Method of Multipliers} (ADMM) method~\cite{boyd2011distributed} is leveraged in our CSB pruning scheme. 
The ADMM separates the weight pruning (during training) problem into two subproblems, which are iteratively solved until convergence.
First, the problem is reformulated as,
{\small
\begin{equation}
\begin{aligned}
  & \minimize_{\{ \mathbf{W}_i\}, \{\mathbf{b}_i\}} && f(\{ \mathbf{W}_i\}_{i=1}^N, \{\mathbf{b}_i\}_{i=1}^N)+\sum_{i=1}^N g_i(\mathbf{Z}_i) \\
  & s.t. && \mathbf{W}_i=\mathbf{Z}_i, i=1,...,N
\end{aligned}
\label{eqn:admm_one}
\end{equation}
}
where $\mathbf{Z}_i$ is an auxiliary variable for subproblem decomposition, and the indicator function (\refeqn{eqn:admm_ind}) is used to replace the original constraint on feasible set.
{\small
\begin{equation}
g_i(\mathbf{Z}_i)= 
\begin{cases}
  0 & \text{if } \mathbf{W}_i \in  \mathbf{S}_i, \\
  +\infty & \text{otherwise.}
\end{cases}
\label{eqn:admm_ind}
\end{equation}
}
Then the \refeqn{eqn:admm_one} can be decomposed to two subproblems listed in \refeqn{eqn:admm_sub1} and \refeqn{eqn:admm_sub2} with the formation of augmented Lagrangian~\cite{fortin2000augmented}. 
{\small
\begin{align}
  &\minimize_{\{ \mathbf{W}_i\}, \{\mathbf{b}_i\}} && f(\{ \mathbf{W}_i\}_{i=1}^N, \{\mathbf{b}_i\}_{i=1}^N)+\sum_{i=1}^N \frac{\uprho_i}{2}||\mathbf{W}_i-\mathbf{Z}_i^t+\mathbf{U}_i^t||^2_F \label{eqn:admm_sub1} \\
  &\minimize_{\{\mathbf{Z}_i\}} && g_i(\mathbf{Z}_i) +\sum_{i=1}^N \frac{\uprho_i}{2}||\mathbf{W}_i^{t+1}-\mathbf{Z}_i+\mathbf{U}_i^t||^2_F \label{eqn:admm_sub2}  
\end{align}
}
where $t$ denotes the iteration index in the ADMM process, $\mathbf{U}_i$ is the dual variable that is updated in each iteration through $\mathbf{U}_i^t=\mathbf{U}_i^{t-1}+\mathbf{W}_i^t-\mathbf{Z}_i^t$.
Following the ADMM process, the two subproblems are iteratively solved till convergence. 
The first subproblem (\refeqn{eqn:admm_sub1}) is solved by the classical SGD method, and the solution for the second subproblem (\refeqn{eqn:admm_sub2}) is obtained by
{\small
\begin{align}
  \mathbf{Z}_i^{t+1}=\proj_{\mathbf{S}_i}(\mathbf{W}_i^{t+1}+\mathbf{U}_i^t)
  \label{eqn:admm_sol}
\end{align}
}
where $\proj$ is the Euclidean projection onto constraint set $\mathbf{S}_i$, which guarantees the weight matrices exhibit the specific sparse pattern defined in the constraint $\mathbf{S}_i$ for each layer. 
In this work, we propose a new type of structured sparse matrix with the novel CSB sparse format, which is the target pattern ($\mathbf{S}_i$) in our RNN weight pruning method. 
The detailed projection process for CSB formated weight will be given in \refsec{sec:csbprune}.


\section{CSB Pruning Technique}

\subsection{A Novel Structured Sparse Weight Format}
\label{sec:csb_def}

\begin{figure}
  \centering
  \includegraphics[width=0.47\textwidth]{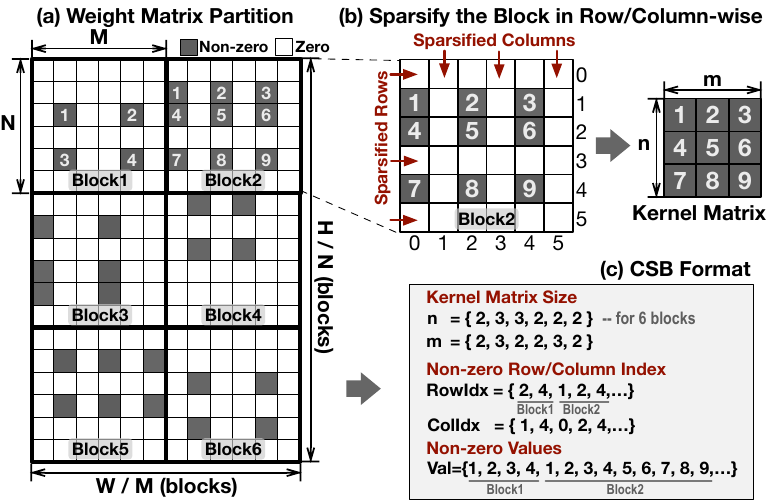}
  \caption{A novel structured sparse matrix (CSB) with its dedicated storage format, which benefits both the pruning flexibility and hardware parallelism.}
  \label{fig:03_csb}
\end{figure}

\subsubsection{Definition of CSB}
We propose the compressed structured block (CSB), a novel \emph{structured sparse} matrix for model pruning that benefits both the pruning flexibility and the hardware parallelism in inference acceleration. 
\reffig{fig:03_csb} illustrates the CSB matrix and the dedicated storage format.
As \reffig{fig:03_csb}(a), we consider the CSB-structured matrix (with a size of $W\times H$) to be composed of multiple \embf{blocks} with the size $M\times N$. 
Each block is sparsified in the \embf{row/column}-wise, as \reffig{fig:03_csb}(b), in which the certain rows/columns are set to zero as a whole. 
By doing so, the non-zero elements are located at the cross-points of the un-sparsified rows/columns only. 
A significant benefit of this \emph{structured sparsity} is the non-zero elements in each block compose a \emph{dense kernel matrix} that provides a higher potential for parallel hardware acceleration than the random sparsity. 
Corresponding to this particular sparsity, a new sparse matrix format is developed for efficient storage and computation.
As \reffig{fig:03_csb}(c), the CSB-format contains five arrays in three groups,
(i) array \hw{n\{\}} and \hw{m\{\}} are the row and column counts of the kernel matrix in each block;
(ii) array \hw{RowIdx\{\}} and \hw{ColIdx\{\}} store the index of un-sparsified (non-zero) rows and columns, respectively;
Note that, the index count for each block equals to the corresponding value in \hw{n\{\}} or \hw{m\{\}};
(iii) the non-zero values in successive blocks (row-major order) are concatenated and stored continuously in the array \hw{Val\{\}}.
Because the inference computation accesses the sparse blocks in sequential, the \emph{offset} for arbitrary access is omitted in the CSB-format. 


\subsubsection{Advantages and Challenges of Pruning with CSB}
We adopt the CSB structured sparsity in pruning the RNN models, which integrates two-fold advantages of both the non-structured pruning and coarse-grained structured pruning in \reffig{fig:02_prune}.
On one hand, CSB provides adequate pruning flexibility, because each block is pruned independently, and the pruning rate varies among blocks that helps to preserve the weights with important information. 
Physically, each element in the weight matrix represents the synapses (connection) between input neurons (matrix column) and output neurons (matrix row). 
The pruning process is zeroing the synapses between two neurons without a strong connection.
The CSB pruning automatically groups the strongly-connected neurons into blocks with high density while leaving the weakly-connected ones in the low-density blocks. 
Further, the \embf{pruning granularity} is adjustable via changing the \emph{block size}; 
Such that different weight matrices in RNN model can be pruned with various granularities. 
The above flexibilities enable a high pruning rate while maintaining the model accuracy. 
On the other hand, the un-pruned weight values in each block compose a dense \emph{kernel matrix} that makes the inference computation friendly to parallel hardware.
Nevertheless, the blocks may have different-sized kernel matrices that result in a \embf{workload imbalance} issue while mapping computation of blocks to parallel hardware.
This paper carefully addresses this issue with an architecture-compilation co-design in \refsec{sec:hw} and \refsec{sec:comp}.


\subsection{CSB Pruning Flow with ADMM}
\label{sec:csbprune}

\begin{algorithm}
\SetKwData{PR}{PruneRate}
\SetKwData{PRStep}{StepPruneRate}
\SetKwData{Flag}{Flag}
\SetKwData{Epoch}{t}
\SetKwData{Up}{up}
\SetKwFunction{Partition}{Partition}
\SetKwFunction{SGD}{SGDTrain}
\SetKwFunction{RowPrune}{RowPrune}
\SetKwFunction{ColumnPrune}{ColumnPrune}
\SetKwFunction{Eval}{Eval}
\SetKwInOut{Input}{input}
\SetKwInOut{Output}{output}

\Input{un-pruned RNN model $\mathbf{W}$; lossless accuracy $accu$, \\
block size in CSB $M\times N$; weight matrix size $W\times H$ \\
initial pruning rate $initPR$ \\
initial step of pruning rate $initPRStep$
}
\Output{ maximally compressed model with CSB pruning $\mathbf{Z}$}

\BlankLine
\tcp{Initialization.}
$\mathbf{U}=\mathbf{0}$; $\mathbf{Z}=\mathbf{W}$; $\mathbf{W}^*=\mathbf{W}$; \Flag=False \\
\PR=$initPR$; \PRStep=$initPRStep$\\

\BlankLine
\tcp{Progressive iteration.}
\Repeat{\PRStep$\leq\frac{1}{4}initPRStep$ \& \Eval{$\mathbf{Z}$}$\geq accu$}{
  \ForEach{\Epoch$\in$ $[0,100)$ \tcp{Re-train and Pruning Epoch.}}{
    \tcp{Solve \refeqn{eqn:admm_sub1} in ADMM (1st subproblem)}
    $\mathbf{W}^*$=\SGD{$\mathbf{W}^*, \mathbf{U}, \mathbf{Z}$} \\
    \BlankLine
    \tcp{Solve \refeqn{eqn:admm_sub2} in ADMM (2nd subproblem)}
    \tcp{Project weight matrix to CSB pattern $\mathbf{S}$}
    $\mathbf{Z}_{i,j}$=\Partition{$\mathbf{W}^*+\mathbf{U}$}, $i\in[0,\frac{W}{M})$, $j\in[0,\frac{H}{N})$ \\
    \ForEach{$j$ $\in$ $[0,H/N)$}{
      $\mathbf{Z}_{:,j}$=\RowPrune{$\mathbf{Z}_{:,j}$, $1-\sqrt{1-\PR}$}
    }
    \ForEach{$i$ $\in$ $[0,W/M)$}{
      $\mathbf{Z}_{i,:}$=\ColumnPrune{$\mathbf{Z}_{i,:}$, $1-\sqrt{1-\PR}$}
    }
    $\mathbf{U}=\mathbf{U}+\mathbf{W}^*-\mathbf{Z}$ \tcp{Update $\mathbf{U}$}
  }
  \BlankLine
  \tcp{Set progressive pruning rate.}
  \If{\Eval{$\mathbf{Z}$}$<accu$}{
    \Flag=True \\
    \PRStep=\PRStep/$2$ \\
    \PR=\PR-\PRStep
  }
  \Else{
    \If{\Flag}{
      \PRStep=\PRStep/$2$
    }
    \PR=\PR+\PRStep
  }  
}
\caption{Auto Lossless CSB Pruning with ADMM}
\label{algo:csbpruning}
\end{algorithm}

With the ADMM-based pruning technique in \refsec{sec:admm_prune}, the weight matrices can be pruned to an arbitrary sparse pattern by defining the constraint $\mathbf{S}$ and applying the pattern-specific projection in \refeqn{eqn:admm_sol}.
To obtain the RNN model with \emph{CSB pattern}, we develop the CSB pruning algorithm following the ADMM principle.
Further, the maximum pruning rate under \emph{lossless} constraint is automatically achieved via the \emph{progressive} pruning.
The entire CSB pruning flow is presented in Algorithm \ref{algo:csbpruning} with carefully specified annotations.
Initially, the baseline model (with dense weight matrix $\mathbf{W}$) is obtained via classical SGD training and input to the flow. 
Note that the bias vector ($\mathbf{b}$) is omitted as the CSB pruning flow does not touch it.
The lossless accuracy ($accu$) is given as the constraint of the progressive pruning.
Two input parameters, initial pruning rate ($initPR$) and initial step of pruning rate reduction ($initPRStep$) are set for tuning the pruning rate in the progressive flow. 
We use the \emph{progressive increase} manner in approaching the maximum value of lossless pruning rate. 
Therefore, we set $initPR$ to a small value (e.g., $4\times$) as the starting point, which surely meets the lossless constraint.
The variables \hw{PruneRate} and \hw{StepPruneRate} are initialized to $initPR$ and $initPRStep$, respectively, at the beginning.
In each progressive iteration, the flow performs re-training and pruning on the model with multiple \emph{epochs} (e.g., $100$ in Algorithm \ref{algo:csbpruning}) to obtain the CSB-formatted weight matrix ($\mathbf{Z}$) with the ADMM-pruning fashion. 
In each epoch, two subproblems are alternatively solved following the principle of the ADMM-pruning technique in \refsec{sec:admm_prune}. 
The function \hw{SGDTrain} updates the weights with classical SGD ($1$st subproblem, \refeqn{eqn:admm_sub1}), and the subsequent process prunes the weight matrix and \emph{projects} it to CSB-constrained set ($2$nd subproblem, \refeqn{eqn:admm_sub2}).
The process in Algorithm \ref{algo:csbpruning} details the \emph{projection} corresponding to the general representation in \refeqn{eqn:admm_sol}.
First, the weight from \hw{SGDTrain} is partitioned to multiple blocks $\mathbf{Z}_{i,j}$ following the CSB method in \refsec{sec:csb_def}.
Then the \hw{RowPrune} process is applied to each block-column independently. 
Specifically, for each block-column, the $\ell_2$-norm (accumulate the square of all elements) of each row (with the size of $M$) is obtained; Then, a row-wise pruning is conducted referring to the $\ell_2$-norm values. 
Subsequently, the \hw{ColumnPrune} is applied to each block-row with the same behavior to \hw{RowPrune}.
Note that the pruning rate in both \hw{RowPrune} and \hw{ColumnPrune} is $1-\sqrt{1-\hw{PruneRate}}$, which results in the target \hw{PruneRate} after the combined processes. 
Once the CSB-formatted weight matrix $\mathbf{Z}$ is obtained, it will be sent to \hw{SGDTrain} of the next epoch, along with un-pruned weight matrix $\mathbf{W}^*$ and accumulated difference matrix $\mathbf{U}$.
With multiple epochs, weight $\mathbf{Z}$ will eventually meet the CSB pattern constraints and achieve good accuracy.

After each progressive iteration, the CSB pruned model is evaluated (\hw{Eval}($\mathbf{Z}$)) and compared to the lossless $accu$.
The \hw{PruneRate} is increased by \hw{StepPruneRate} in the next iteration if the $accu$ is achieved. 
Once \hw{Eval}$(\mathbf{Z})<accu$, the model is over-pruned and the optimum pruning rate is just between the \hw{PruneRate} of the two neighboring iterations.
Therefore, we reduce \hw{StepPruneRate} by half and reduce the \hw{PruneRate} by this new step to further approach the optimum point.
The progressive CSB pruning flow terminates until the pruning rate reaches a target precision.
For instance, as the last line in Algorithm \ref{algo:csbpruning}, the flow terminates when the pruning rate precision (\hw{StepPruneRate}) $\leq$ $\frac{1}{4}initPRStep$.

\section{Unified Architecture for CSB-RNN}
\label{sec:hw}
\subsection{Overview of Acceleration Framework}
\begin{figure}
  \centering
  \includegraphics[width=0.5\textwidth]{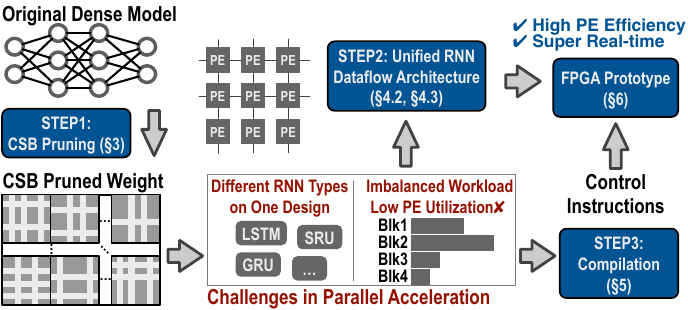}

  \caption{Overview of CSB RNN acceleration framework, including (i) CSB pruning algorithm, (ii) unified RNN dataflow architecture, (iii) workload compilation with CSB pruned model.}
  \label{fig:04_bigpic}
\end{figure}

An overview of the CSB-RNN acceleration framework is illustrated in \reffig{fig:04_bigpic}.
Although the CSB pruning (\hw{STEP1}) shrinks the model size and therefore reduces the computation in inference, 
parallel hardware acceleration is still in demand to achieve realtime performance.
The challenges in accelerating CSB-RNN are two-fold.
First, the architecture should be adaptive to various RNN cell types, i.e., LSTM, GRU, etc. 
Second, the kernel matrix in fine-grained blocks may not provide enough \embf{inner-block} parallelism for large-scale hardware.
To further improve the concurrency, \embf{inter-block} parallelism should be leveraged. 
However, the pruned blocks may have different sparsities, leading to the \embf{workload imbalance issue} for inter-block parallelism, which usually causes a low utilization of processing element (PE).
To address these challenges, CSB-RNN proposes an architecture-compilation co-design.
In the architecture aspect (\hw{STEP2}), we propose a unified RNN dataflow architecture that is programmable for different RNN cell types (\refsec{sec:hw_1}); 
In particular, a novel \csbengine is designed with the support of \embf{workload sharing} and is equipped in CSB-RNN architecture to address the workload imbalance issue (\refsec{sec:hw_2}).
In the compilation aspect (\hw{STEP3}), we define control instructions for the hardware and propose the compilation algorithms to conduct the particular RNN type computation and balanced workload scheduling (\refsec{sec:comp}).

\subsection{Programmable RNN Dataflow Architecture}
\label{sec:hw_1}

\begin{figure}
  \centering
  \includegraphics[width=0.45\textwidth]{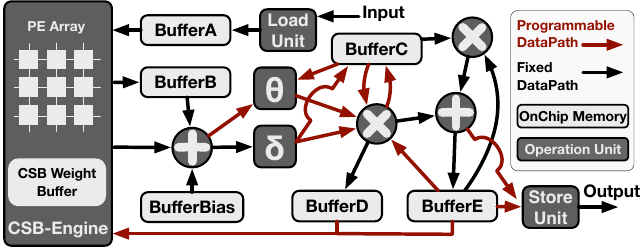}

  \caption{RNN dataflow architecture. Operation units serve the RNN arithmetic primitives; The programmable datapaths construct the proper dataflow for target RNN cell via instructions.}
  \label{fig:04_arch}
\end{figure}

To generalize the architecture for different RNN cell types, we investigated the existing RNN cells and extracted the \embf{arithmetic primitives}, which compose the RNN computation in a \emph{dataflow} fashion.
\reffig{fig:04_arch} presents the hardware components in this architecture, where each operation unit serves the corresponding arithmetic primitive. 
In particular, the \csbengine computes the main workload, MVM, with the weight matrices after CSB pruning (CSB-MVM).
The units \hw{\times}, \hw{+} are the element-wise multiplication and addition. 
\hw{\delta} and \hw{\theta} operate the activation functions \emph{Sigmoid} and \emph{Tanh}, respectively. 
The \emph{datapaths} (arrows on \reffig{fig:04_arch}) interconnect the operation units and on-chip buffers, which transmit the intermediate results and compose the dataflow graph for RNN cell computation.
Importantly, RNN dataflow architecture provides the \embf{programmable datapath} (red arrows on \reffig{fig:04_arch}).
Thus, the proper operation units can be interconnected by programming control instructions for a particular RNN cell type. 

\subsection{CSB-Engine}
\label{sec:hw_2}
The CSB pruning scheme greatly shrinks the weight matrix size and therefore reduces the main workload in inference.
Although the fine-grained structure of CSB contributes to the regularity and makes efficient hardware acceleration possible, it is still challenging to design a parallel architecture that can fully exploit the benefits of CSB pruning.
The challenges in an efficient \csbengine design are two-fold.
First, both the \emph{inner-block} and \emph{inter-block} parallelism should be fully exploited, as the regular inner-block computation provides very limited concurrency with small block size.
Second, the \embf{inter-block workload imbalance issue} exists due to the sparsity varies among blocks.
The following \refsec{sec:04_basiccsb} and \refsec{sec:04_sharing} address these two challenges, respectively.

\subsubsection{Hierarchical Design for Inner- and Inter-Block Parallelism}
\label{sec:04_basiccsb}

\begin{figure}
  \centering
  \includegraphics[width=0.5\textwidth]{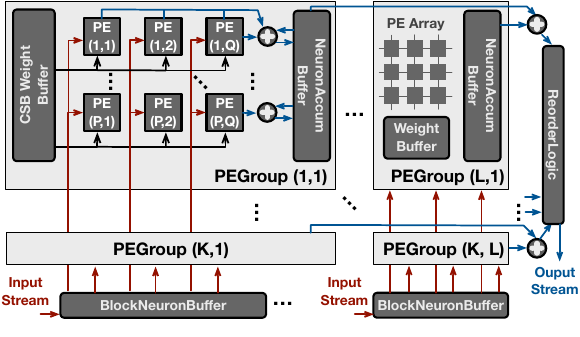}
  \caption{Two-level hierarchical organization of CSB-Engine for the main workload (CSB-MVM) computation.}
  \label{fig:04_csbengine}
\end{figure}

As illustrated in \reffig{fig:04_csbengine}, the \csbengine design is in a two-level hierarchy, processing element (\hw{PE}) level and \hw{PEGroup} level.
The hardware instances in each level are organized in a 2D fashion that the architecture is composed of $K\times L$ \hw{PEGroup}s, and each \hw{PEGroup} contains $P\times Q$ \hw{PE}s. 
The parallel \hw{PE}s inside one \hw{PEGroup} process inner-block multiplication concurrently, while the \hw{PEGroup}s computing different blocks in parallel (inter-block parallelism).

Inside each \hw{PEGroup}, because the size of CSB kernel matrix ($m\times n$) might be larger than that of \hw{PE} array ($P\times Q$), multi-pass processing is required to handle the entire block. 
Thus, each \hw{PEGroup} contains a \hw{NeuronAccumBuffer}, which stores the partial results and sums up with the accumulation of horizontal \hw{PE}s in each pass. 
The input neurons required by the current block are preloaded to the \hw{BlockNeuronBuffer} and broadcasted to the \hw{PE} array. 
Each \hw{PE} column shares the same input neuron as the unpruned weights are vertically aligned in the structured block with CSB pruning. 
Importantly, the \hw{WeightBuffer} provides the CSB-formatted weight (\reffig{fig:03_csb}), including the weight values (kernel matrix) for \hw{PE}s, column index for \hw{BlockNeuronBuffer} to read the proper input neuron, row index for \hw{NeuronAccumBuffer} to accumulate the multiplication-results to proper address in \hw{NeuronAccumBuffer}, and the kernel matrix size ($m\times n$) for the \hw{PEGroup} control logic which conducts proper pass count in both axes.

In the higher-level of the design hierarchy, the \hw{PEGroup}s process blocks in the row-major order. 
The \hw{PEGroup}s in one column concurrently compute the vertical blocks.
Therefore, the \hw{PEGroup}s in one column share the same partition of input neuron vector, while multi-ports are provided on \hw{BlockNeuronBuffer} for concurrent access.
Similarly, the blocks in horizontal axis are mapped to \hw{PEGroup}s in the same row, with multi-pass processing. 
After the computation of each block-row, the results in \hw{NeuronAccumBuffer}s are accumulated in horizontal and output to \hw{ReorderLogic} to obtain the output neuron vector.

\subsubsection{Inter-PEGroup Workload Sharing}
\label{sec:04_sharing}

\begin{figure*}
  \centering
  \includegraphics[width=0.95\textwidth]{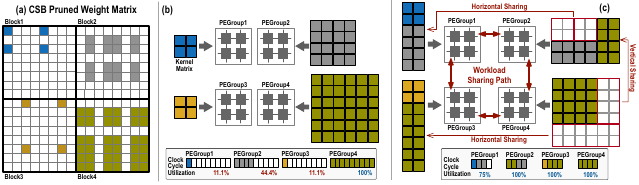}
  \caption{Inter-block workload imbalance issue occurs when mapping the CSB pruned matrix (a) to the vanilla (basic) CSB-Engine (b), which results in a low hardware utilization. We propose the workload sharing technique that significantly increases the utilization and reduces the time consumption, as demonstrated in (c).}
  \label{fig:04_imbalance}
\end{figure*}

~

\textbf{Workload Imbalance Challenge:}
The blocks in CSB pruned model may have different-sized kernel matrices, and the resultant \emph{inter-block workload imbalance} brings challenges to exploit the inter-block parallelism on hardware. 
As \reffig{fig:04_imbalance}(b) demonstrates, with the \emph{straightforward} design, the workload imbalance issue results in low utilization of \hw{PEGroup}s. 
The presented MVM workloads are allocated to $2\times 2$ \hw{PEGrounp}s that each contains $4$ \hw{PE}s.
During the execution, \hw{PEGroup1}-\hw{3} enter the idle state before the \hw{PEGroup4} accomplishes its workload, which results in a severe under-utilization of the overall hardware. 
In fact, the imbalanced sparsity naturally exists in the RNN models.
However, existing works~\cite{han2017ese, cao2019efficient} relieve the hardware pain by pruning the model with a region-balanced sparsity compulsively. 
As a result, the neglect of natural sparsity-imbalance significantly harms the pruning ratio and model accuracy. 
By contrast, we handle this issue by improving the architecture with the \embf{workload sharing} technique.

\textbf{Inter-PEGroup Workload Sharing:}
The concept of workload sharing is illustrated in \reffig{fig:04_imbalance}(c). 
Each \hw{PEGroup} processes not only the originally allocated block but also a partition of block from the neighboring \hw{PEGroup}, which is arranged with a heavier workload. 
In the hardware aspect, as \reffig{fig:04_imbalance}(c), dedicated workload sharing paths (red arrows) are set for the inter-\hw{PEGroup} data transmission, and the interconnection adopts the \emph{torus topology} in both dimensions. 
With the hardware support of workload sharing, \hw{PEGroup4} migrates the extra workloads to \hw{PEGroup2} and \hw{PEGroup3}; And \hw{PEGroup2} migrates the Block$2$ workload partition to \hw{PEGroup1}. That significantly balances the workload and improves the utilization.
Considerations in the workload sharing design are two-fold. 
(i) The input neurons should be sharable between the \hw{PEGroup}s;
(ii) The output neuron accumulation should be performed inter-\hw{PEGroup}s.
We discuss these issues and our strategies within two cases, in which the workload is shared between neighboring \hw{PEGroup}s in horizontal or in vertical, respectively. 
For the \emph{horizontal sharing} case, an extra data port is set on the \hw{BlockNeuronBuffer} to solve the issue (i), which enables the \hw{PEGroup} to access input neurons from the neighboring \hw{PEGroup} in horizontal.
The issue (ii) is naturally solved by the hierarchical \csbengine design, as the \hw{PEGroup} can store the partial results of the shared workload partition in its local \hw{NeuronAccumBuffer}, which will be accumulated in horizontal after processing the entire block-row.
For the \emph{vertical sharing} case, the \hw{PEGroup}-column shares the same \hw{BlockNeuronBuffer}, thus the issue (i) is naturally solved by hardware. 
About the issue (ii), the \hw{PEGroup} should accumulate the vertically shared workload to its original \hw{PEGroup}, as the vertical \hw{PEGroup}s compute different block-rows that cannot be accumulated in a mixed manner.
However, concurrent accumulation to one address in \hw{NeuronAccumBuffer} leads to the read-after-write (RAW) data hazard.
To address this issue, an accumulation path is set between vertical \hw{PEGroup}s and connected to the adder, which accepts parallel results from neighboring \hw{PEGroup}s, sums up and stores to the \hw{NeuronAccumBuffer} for one shot. 
With the hardware support on workload sharing, we propose the compilation scheme in next section that schedules the partition and sharing by analyzing the CSB pruned matrix and generates the instruction to control the hardware-sharing behavior.

\section{Compilation for CSB Pruned Model}
\label{sec:comp}
The proposed RNN dataflow architecture is controlled by the pre-compiled instructions.
The instruction set includes the \embf{macro-instruction} and \embf{micro-instruction}, where the former one conducts the operation units (in \reffig{fig:04_arch}) for the proper RNN dataflow (cell type); 
and the later one instructs the \csbengine with inter-\hw{PEGroup} workload sharing behavior as described in \refsec{sec:04_sharing}.
Correspondingly, the compilation is composed of two phases, \emph{RNN dataflow} compilation (\refsec{sec:comp_macro}) and workload sharing scheduling (\refsec{sec:comp_micro}).


\subsection{RNN Cell to Dataflow Architecture}
\label{sec:comp_macro}
\subsubsection{Macro-Instruction Set} 
\begin{figure}
  \centering
  \includegraphics[width=0.48\textwidth]{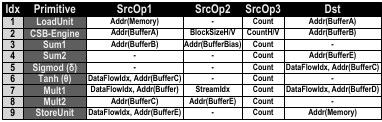}
  \caption{Macro-instruction set (VLIW-like) for RNN dataflow architecture that constructs proper arithmetic dataflow for different RNN cell types.}
  \label{fig:05_macroinst}
\end{figure}

We define the macro-instruction set for our RNN dataflow architecture (\refsec{sec:hw_1}).
As \reffig{fig:05_macroinst}, the micro-instruction is composed of multiple sections, that each section provides control signals for corresponding RNN primitive hardware.
All sections are concatenated to form a very long instruction word (VLIW) item. 
Note that each section contains \hw{Count} operand to indicate the \emph{size of workload} for corresponding hardware primitive. 
Thus, one VLIW instruction is regarded as accomplished until all hardware primitives finish the workload. 
The operands in each instruction section are classified into two types, the \hw{Count} type controls the hardware iteration count, and the other operands indicate the proper \emph{data source or destination} for each primitive. 
For the first type, the value of \hw{Count} in element-wise operation units (only \csbengine excluded) is measured by data element as these units perform \emph{element-wise} operation. 
Differently, the \hw{CountH/V} in \csbengine section represents the horizontal/vertical block iteration counts over the entire \csbengine in processing the particular weight matrix.
For the second operand type, \hw{Addr(Memory)} and \hw{Addr(Buffer)} give the access address of external memory (normally DRAM) and built-in buffers in the architecture, respectively. 
Importantly, the programmable datapaths in the architecture (\reffig{fig:04_arch}) are indexed, and the \hw{DataFlowIdx} is set in the operand to indicate the proper data source or destination for hardware primitive. 
With the above settings, RNN models with various cell types can be translated to several VLIW instructions that are repetitively executed during RNN inference. 

\subsubsection{Macro-Instruction Compilation}
The objective of compilation is to minimize the VLIW instruction count that maximizes the utilization of operation units.
We invoke the \emph{list scheduling} method~\cite{lam1988software} that is usually applied in VLIW compilation.
The RNN model with a particular cell type is translated to the \emph{directed acyclic graph} (DAG), in which the nodes represent the arithmetic primitives and the edges are data dependencies.
In the list scheduling, we adopt the as soon as possible (ASAP) strategy that the operation nodes are mapped to the corresponding hardware resources once the resource is available and the dependency edge is ready. 
With the proper operation units and interconnection in the RNN dataflow architecture, the macro-instruction compilation can quickly achieve an optimum point, in which the processing throughput is bounded by the main workload (CSB-MVM) on \hw{CSB}-\hw{Engine}.

\subsection{Workload Scheduling on CSB-Engine}
\label{sec:comp_micro}

\subsubsection{Micro-Instruction Set}

\begin{figure}
  \centering
  \includegraphics[width=0.5\textwidth]{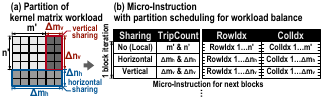}
  \caption{Micro-instruction indicates the kernel matrix workload and the scheduling of partition for workload balancing.}
  \label{fig:05_microinst}
\end{figure}


The micro-instructions are generated for each \hw{PEGroup} individually, which control the CSB-MVM operations on \hw{CSB}-\hw{Engine}. 
Specifically, the micro-instruction contains the CSB-compression information (i.e., kernel matrix size, row- and column-index in \reffig{fig:03_csb}(c)) for the block workload allocated to the certain \hw{PEGroup}.
In particular, the kernel matrix workload is partitioned to \emph{three submatrices} and shared to neighboring \hw{PEGroup}s (as \reffig{fig:05_microinst}(a)), the micro-instructions for one block iteration include three items,
(i) \emph{local} workload that is originally allocated, excluding the portion shared to other \hw{PEGroup}s; (ii) workload shared from the neighboring \hw{PEGroup} in \emph{horizontal}; (iii) workload shared from the neighboring \hw{PEGroup} in \emph{vertical}. 
The micro-instruction contains $4$ operands, as \reffig{fig:05_microinst}(b).
The operand \hw{Sharing} gives a flag (local/horizontal/vertical) to indicate the data source, where \emph{local} means the input and output neurons are in local \hw{PEGroup}; 
\emph{horizontal} (sharing) indicates the input neurons should be read from the \hw{BlockNeuronBuffer} of left \hw{PEGroup}; 
\emph{Vertical} (sharing) means the output should be accumulated to the \hw{NeuronAccumBuffer} of upper \hw{PEGroup}.
The operand \hw{TripCount} gives the size of workload. 
Note that, for each block, the kernel matrix is divided to tree regular partitions as \reffig{fig:05_microinst}(a), for local (no-sharing), vertical- and horizontal-sharing, respectively. 
The sizes of partitioned matrices are denoted as $m^\prime \times n^\prime$, $\Delta m_v \times \Delta n_v$, $\Delta m_h \times \Delta n_h$, which are turned to \hw{TripCount} values in the three micro-instruction items.
The operands \hw{RowIdx} and \hw{ColIdx} provide the non-zero row and column indices of each submatrix. 
Note that each micro-instruction item may contain multiple \hw{RowIdx} and \hw{ColIdx} corresponding to the \hw{TripCount} value. 
Further, these two operands are stored in individual instruction memories that are read and reused periodically in the submatrix computation.

\subsubsection{Micro-Instruction Compilation}
The compilation of micro-instruction is essentially searching the workload partition scheme to achieve the optimal balance, which facilitates a higher hardware utilization (efficiency).
Specifically, the compiler analyzes the weight matrices and selects the proper partition variables (as \reffig{fig:05_microinst}(a)) for each kernel matrix.
Every $K\times L$ blocks (one block iteration) are analyzed individually, which are executed on \hw{PEGroup}s in concurrent. 
Within one block iteration, each \hw{PEGroup} is supposed to take the equivalent workload after balancing. 

We regard the search of optimal partition variable as a satisfiability modulo theories (SMT) problem~\cite{winter1992path}, which searches the feasible solutions in the constrained region. 
The existing SMT solver~\cite{de2008z3} takes the constraints with logic programming and gives the satisfiability (existence of solution) and feasible solutions.
In the compilation for each block iteration, we declare the integer variables including $m^\prime(k,l)$, $n^\prime(k,l)$, $\Delta m_h(k,l)$, $\Delta n_h(k,l)$, $\Delta m_v(k,l)$, $\Delta n_v(k,l)$, where $k\in [1,K]$ and $l\in [1,L]$.
The constraints are represented with the constraint logic programming (CLP), in which each \emph{clause} gives a specific search limitation.
The CLP in compilation is listed in \refeqn{eqn:clp}, where $\wedge$ represents logic AND and $\vee$ represents OR.
\hws{CLP}{1,2} constraint the feasible search region, as the size of the partitioned workload should $\leq$ kernel matrix size ($m\times n$).
\hws{CLP}{3,4} guarantee regular partitions as \reffig{fig:05_microinst}(a). 
\hws{CLP}{5} determines the values of $m^\prime$ and $n^\prime$.
To improve the \hw{PEGroup} utilization, we set \hws{CLP}{6} constraint that the size of partition workload should be integer-multiple of the \hw{PEGroup} size. 
Thus, the \hw{PE}s are fully utilized on the shared workload partition. 
Also, it helps to shrink the search space and speed up the compilation.
Within the idealized situation, each \hw{PEGroup} is scheduled with workload that is the average value over all \hw{PEGroup}s in the current block iteration.
Otherwise, the \hw{PEGroup} with \emph{maximum} workload determines the run time (clock cycle) for this iteration.
\hws{CLP}{7} gives the constraint on the \emph{maximum} workload that, to all \hw{PEGroup}s, the exceeding part of scheduled workload to the average value ($avg$) should $\leq$ $margin$, which is given before search.
The last \hw{CLP} combines all above constraints to a conjunctive form, which is subsequently sent to SMT-solver for a feasible solution. 

{\footnotesize
\begin{align}
  &\hws{CLP}{1}: &&\{0 \leq \Delta m_h(k,l)\leq m(k,l)\} \wedge \{0 \leq \Delta n_h(k,l)\leq n(k,l)\} \nonumber \\
  &\hws{CLP}{2}: &&\{0 \leq \Delta m_v(k,l)\leq \lfloor m(k,l)/2\rfloor \} \wedge \{0 \leq \Delta n_v(k,l)\leq n(k,l)\} \nonumber \\
  &\hws{CLP}{3}: &&\{\Delta m_h(k,l)=m(k,l)\} \wedge \{\Delta n_v(k,l)+\Delta n_h(k,l)=n(k,l) \} \nonumber \\
  &\hws{CLP}{4}: &&\{\Delta n_v(k,l)=n(k,l)\} \wedge \{\Delta m_h(k,l)+\Delta m_v(k,l)=m(k,l) \} \nonumber \\
  &\hws{CLP}{5}: &&\{m^\prime(k,l)=m(k,l)-\Delta m_v(k,l)\} \wedge \{n^\prime(k,l)=n(k,l)-\Delta n_h(k,l)\} \nonumber \\  
  &\hws{CLP}{6}: &&\{\Delta m_h(k,l)\%P=m_v^\prime(k,l)\%P=0 \} \wedge \{\Delta n_h(k,l)\%Q=n_v^\prime(k,l)\%Q=0\} \nonumber  \\
  &\hws{CLP}{7}: && |~(m^\prime(k,l)\times n^\prime(k,l)+ \Delta m_h(k,l-1)\times \Delta n_h(k,l-1) \nonumber \\
  & && + \Delta m_v(k-1,l)\times \Delta n_v(k-1,l)) - avg~| \leq margin \nonumber \\
  &\hw{CLP}: && \hws{CLP}{1} \wedge \hws{CLP}{2} \wedge (\hws{CLP}{3} \vee \hws{CLP}{4}) \wedge \hws{CLP}{5} \wedge \hws{CLP}{6} \wedge \hws{CLP}{7}
  \label{eqn:clp}
\end{align}
}

\begin{algorithm}
\SetKwData{PR}{PruneRate}
\SetKwData{PRStep}{StepPruneRate}
\SetKwData{PartitionVar}{PartitionVar}
\SetKwData{Sat}{Satisfiability}
\SetKwData{CLP}{CLP}

\SetKwFunction{SMTSolver}{SMTSolver}
\SetKwFunction{BuildCLP}{BuildCLP}
\SetKwFunction{Append}{Append}
\SetKwFunction{ColumnPrune}{ColumnPrune}
\SetKwFunction{Analyze}{Analyze}

\SetKwInOut{Input}{input}
\SetKwInOut{Output}{output}

\Input{CSB pruned weight matrix $\mathbf{W}_{csb}$; \\
block size in CSB $M\times N$; weight matrix size $W\times H$; \\
size of each \hw{PEGroup} $P\times Q$; \hw{PRGroup} count $K\times L$
}
\Output{Micro-instruction list $MicroInst$}

\tcp{Temporal block iterations in vertical.}
\For{$i\leftarrow 1$ \KwTo $\lceil H/N/K \rceil$ }{
\tcp{Temporal block iterations in horizontal.}
\For{$j\leftarrow 1$ \KwTo $\lceil W/M/L \rceil$ }{
\BlankLine
$margin$=0 \\
\tcp{$\forall k\in [1,K]$, $\forall l\in [1,L]$.}
[$m(k,l), n(k,l)$, $avg$]=\Analyze($\mathbf{W}_{csb}$,i,j) \\
\BlankLine
\tcp{Search with multiple rounds.}
\Repeat{\Sat=True}{
  \CLP=\BuildCLP($m(k,l), n(k,l)$, $avg$, $margin$) \\
  \tcp{Give solution if satisified.}
  [\Sat, \PartitionVar] = \SMTSolver(\CLP) \\
  $margin$+=$P\times Q$
}
$MicroInst$=\Append($MicroInst$, \PartitionVar)
}}
\caption{Micro-Instruction Compilation}
\label{algo:compile}
\end{algorithm}
Based on the above formulation, we propose the compilation scheme in Algorithm \ref{algo:compile} that seeks out the optimal scheduling solution.
For a given CSB formatted weight matrix $\mathbf{W}_{csb}$, the compiler partitions it to $\lceil W/M/L \rceil \times \lceil H/N/K \rceil$ \emph{temporal block iterations} and schedules each iteration individually. 
Before the multi-round search, the compiler firstly analyzes the weight partition for current block iteration that gives the kernel matrix size ($m,n$) for each block and the average workload ($avg$).
The $margin$ is initialized to $0$ that targets to schedule an idealized average workload on each \hw{PEGroup}.
In the search round, \hw{BuildCLP} constructs the constraints representation, which is input to \hw{SMTSolver}.
In case the constraints cannot be satisfied (\hw{Satisfiability} is False) over the feasible region, the $margin$ value is supposed to increase by $P\times Q$ in the next round. 
Once the SMT problem is satisfied, the search stops and the partition variables ($m^\prime$, $n^\prime$, $\Delta m_v$, $\Delta n_v$, $\Delta m_h$, $\Delta n_h$) for each \hw{PEGroup} are assembled and appended to the micro-instruction list, that conducts the \csbengine computation in a workload balanced fashion.

\section{Evaluation}
In this section, we first brief the implementation of the CSB-RNN framework (\refsec{sec:eval_impl}), and then give deep evaluations from the performance of CSB pruning algorithm (\refsec{sec:eval_rate}) to the improvement with the architecture-compilation co-design (\refsec{sec:eval_hw}).
Meanwhile, $10$ mainstream RNN models from multi-domains are invoked as the evaluation benchmarks and presented in \reftab{tbl:benchmark}, in which we also list the non-structured pruning rates as the theoretical optimum.

\subsection{Implementation and Experiments Setup}
\label{sec:eval_impl}
The CSB pruning flow was implemented with PyTorch~\cite{paszke2019pytorch}, a framework for deep learning model development. 
The benchmark models were first trained with the SGD and the accuracy is regarded as the \emph{lossless target value} in the subsequent CSB pruning. 
These baseline models were fed in the CSB pruning flow and get compressed with the lossless constraints. 
In regarding the architecture-compilation co-design, the proposed RNN dataflow architecture was realized with Verilog RTL and implemented on an FPGA vendor evaluation board (Xilinx-ZCU$102$), on which the FPGA contains enough resources for our architecture with different design scales. 
The compiler was implemented in C++ with the strategies in \refsec{sec:comp} and Z3~\cite{de2008z3} as the SMT solver. 
With the CSB pruned model, the compiler dumps the macro-instructions (\refsec{sec:comp_macro}) to build the proper RNN dataflow and micro-instructions (\refsec{sec:comp_micro}) for the workload balancing. 
These instructions are loaded to the RNN dataflow architecture before processing sequence continuously. 
With regard to the detailed hardware efficiency (i.e, \csbengine utilization), cycle-level RTL simulation was performed to profile the inference behavior. 

\begin{table*}[tb]
\footnotesize
\ra{1.05}
\setlength\tabcolsep{1.6pt}
\centering
\caption{Benchmark Models in CSB-RNN Evaluation}
\vspace{-1em}
\begin{tabular}{cll|c|clclll|ll|cll}
\toprule
\multicolumn{1}{l}{\textbf{App.}} & \multirow{2}{*}{\textbf{Abbr.}} & \multicolumn{1}{c|}{\multirow{2}{*}{\textbf{Applications}}} & \multirow{2}{*}{\textbf{Dataset}} & \multirow{2}{*}{\textbf{\#Layer}} & \multirow{2}{*}{\textbf{RNN Cell}} & \textbf{Layer} & \multicolumn{1}{c}{\textbf{\#Input}} & \multicolumn{1}{c}{\textbf{\#Hidden}} & \multirow{2}{*}{\textbf{Evaluation Metric}} & \multicolumn{2}{c|}{\textbf{Original Model}} & \multicolumn{3}{c}{\textbf{Non-Structued Pruning}} \\
\multicolumn{1}{l}{\textbf{Idx.}} &  & \multicolumn{1}{c|}{} &  &  &  & \textbf{Index} & \multicolumn{1}{c}{\textbf{Neuron}} & \multicolumn{1}{c}{\textbf{Neuron}} &  & \multicolumn{1}{c}{\#Weight+Bias} & \multicolumn{1}{c|}{Result} & PruneRate & \multicolumn{1}{c}{\#Weight} & \multicolumn{1}{c}{Result} \\ \hline
\multirow{2}{*}{1} & \multirow{2}{*}{\hw{MT1}} & \multirow{2}{*}{Machine Translation} & \multirow{2}{*}{PTB\cite{marcus1993building}} & \multirow{2}{*}{2} & \multirow{2}{*}{LSTM\cite{hochreiter1997long}} & 1 & 128 & 256 & \multirow{2}{*}{\begin{tabular}[c]{@{}l@{}}Perplexity\\ (PPL, lower is better)\end{tabular}} & 393K+1K & \multirow{2}{*}{110.89} & 13.2$\times$ & 29.8K & \multirow{2}{*}{111.62} \\
 &  &  &  &  &  & 2 & 256 & 256 &  & 524K+1K &  & 13.2$\times$ & 39.7K &  \\ \hline
\multirow{2}{*}{2} & \multirow{2}{*}{\hw{MT2}} & \multirow{2}{*}{Machine Translation} & \multirow{2}{*}{PTB\cite{marcus1993building}} & \multirow{2}{*}{2} & \multirow{2}{*}{LSTM\cite{hochreiter1997long}} & 3 & 1500 & 1500 & \multirow{2}{*}{Perplexity (PPL)} & 18M+6K & \multirow{2}{*}{80.66} & 16.3$\times$ & 1.1M & \multirow{2}{*}{82.33} \\
 &  &  &  &  &  & 4 & 1500 & 1500 &  & 18M+6K &  & 16.3$\times$ & 1.1M &  \\ \hline
\multirow{2}{*}{3} & \multirow{2}{*}{\hw{SR1}} & \multirow{2}{*}{Speech Recognition} & \multirow{2}{*}{TIMIT\cite{garofolo1993darpa}} & \multirow{2}{*}{2} & \multirow{2}{*}{LSTMP\cite{sak2014long}} & 5 & 153 & 1024 & \multirow{2}{*}{\begin{tabular}[c]{@{}l@{}}Phoneme Error Rate\\ (PER, lower is better)\end{tabular}} & 3.25M+4K & \multirow{2}{*}{19.39\%} & 14.5$\times$ & 224.0K & \multirow{2}{*}{19.70\%} \\
 &  &  &  &  &  & 6 & 512 & 1024 &  & 4.72M+4K &  & 14.5$\times$ & 325.4K &  \\ \hline
\multirow{2}{*}{4} & \multirow{2}{*}{\hw{SR2}} & \multirow{2}{*}{Speech Recognition} & \multirow{2}{*}{TIMIT\cite{garofolo1993darpa}} & \multirow{2}{*}{2} & \multirow{2}{*}{GRU\cite{cho2014learning}} & 7 & 39 & 1024 & \multirow{2}{*}{PER} & 3.3M+3K & \multirow{2}{*}{19.24\%} & 21.7$\times$ & 150.5K & \multirow{2}{*}{19.80\%} \\
 &  &  &  &  &  & 8 & 1024 & 1024 &  & 6.3M+3K &  & 21.7$\times$ & 289.9K &  \\ \hline
\multirow{2}{*}{5} & \multirow{2}{*}{\hw{SR3}} & \multirow{2}{*}{Speech Recognition} & \multirow{2}{*}{TIMIT\cite{garofolo1993darpa}} & \multirow{2}{*}{2} & \multirow{2}{*}{Li-GRU\cite{ravanelli2018light}} & 9 & 39 & 512 & \multirow{2}{*}{PER} & 564.2K & \multirow{2}{*}{16.87\%} & 7.1$\times$ & 79.5K & \multirow{2}{*}{17.30\%} \\
 &  &  &  &  &  & 10 & 512 & 512 &  & 1M &  & 7.1$\times$ & 147.7K &  \\ \hline
6 & \hw{SR4} & Speech Recognition & TDIGIT\cite{tdigit} & 1 & GRU\cite{cho2014learning} & 11 & 39 & 256 & Accuracy & 226.6K+0.8K & 99.98\% & 25.7$\times$ & 8.8K & 99.21\% \\ \hline
\multirow{2}{*}{7} & \multirow{2}{*}{\hw{SPP}} & \multirow{2}{*}{Stock Price Prediction} & \multirow{2}{*}{S\&P500\cite{sp500}} & \multirow{2}{*}{2} & \multirow{2}{*}{LSTM\cite{hochreiter1997long}} & 12 & 1 & 128 & \multirow{2}{*}{\begin{tabular}[c]{@{}l@{}}Normalized Price Dist.\\ (lower is better)\end{tabular}} & 66K+0.5K & \multirow{2}{*}{0.47} & 4.1$\times$ & 16.1K & \multirow{2}{*}{0.51} \\
 &  &  &  &  &  & 13 & 128 & 128 &  & 131K+0.5K &  & 4.1$\times$ & 32K &  \\ \hline
\multirow{3}{*}{8} & \multirow{3}{*}{\hw{SC1}} & \multirow{3}{*}{Sentiment Classification} & \multirow{3}{*}{IMDB\cite{maas2011learning}} & \multirow{3}{*}{3} & \multirow{3}{*}{LSTM\cite{hochreiter1997long}} & 14 & 32 & 512 & \multirow{3}{*}{\begin{tabular}[c]{@{}l@{}}Accuracy\\ (higher is better)\end{tabular}} & 1.11M+2K & \multirow{3}{*}{86.37\%} & 10.4$\times$ & 107.1K & \multirow{3}{*}{85.65\%} \\
 &  &  &  &  &  & 15 & 512 & 512 &  & 2.1M+2K &  & 10.4$\times$ & 201.6K &  \\
 &  &  &  &  &  & 16 & 512 & 512 &  & 2.1M+2K &  & 10.4$\times$ & 201.6K &  \\ \hline
9 & \hw{SC2} & Sentiment Classification & MR\cite{pang2005seeing} & 1 & LSTM\cite{hochreiter1997long} & 17 & 50 & 256 & Accuracy & 313.3K+1K & 78.23\% & 7.2$\times$ & 43.5K & 76.31\% \\ \hline
\multirow{3}{*}{10} & \multirow{3}{*}{\hw{QA}} & \multirow{3}{*}{Question Answering} & \multirow{3}{*}{BABI\cite{weston2015towards}} & \multirow{3}{*}{3} & \multirow{3}{*}{LSTM\cite{hochreiter1997long}} & 18 & 50 & 256 & \multirow{3}{*}{Accuracy} & 313.3K+1K & \multirow{3}{*}{65.37\%} & 7.9$\times$ & 39.7K & \multirow{3}{*}{64.51\%} \\
 &  &  &  &  &  & 19 & 256 & 256 &  & 524.3K+1K &  & 7.9$\times$ & 66.4K &  \\
 &  &  &  &  &  & 20 & 256 & 256 &  & 524.3K+1K &  & 7.9$\times$ & 66.4K &  \\ \hline
\bottomrule
\end{tabular}
\label{tbl:benchmark}
\end{table*}

\subsection{Evaluation of CSB pruning Rate}
\label{sec:eval_rate}

\begin{figure*}
  \centering
  \includegraphics[width=1\textwidth]{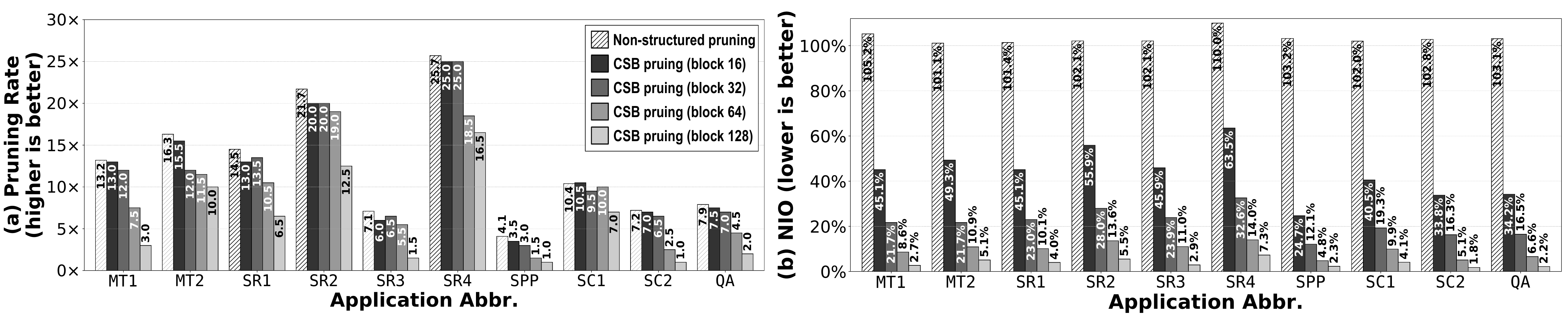}
  \vspace{-2em}
  \caption{(a) shows the pruning rate comparison between non-structured pruning (optimum) and CSB pruning in different block sizes. (b) shows the normalized index overhead (NIO). Comparing (a) and (b), we gain the insight that CSB pruning dramatically reduces the NIO while maintaining a high pruning rate.}
  \label{fig:06_pr1}
\end{figure*}

The CSB pruning is first evaluated in the aspect of \emph{pruning rate}, which is a significant metric to score the model compression methods.
Because the parameterizable block size determines the structural granularity in pruning, we present the attainable maximum pruning rate with various block sizes in \refsec{sec:eval_rate_1}. 
Further, comparison with the prior art RNN compression schemes is given in \refsec{sec:eval_rate_2}.

\subsubsection{Selection of Optimum Structural Granularity}
\label{sec:eval_rate_1}
CSB pruning provides the \emph{flexibility} that improves the pruning rate and also the hardware-friendly \emph{regularity}. 
Importantly, a trade-off exists between these two targets that motivate the following investigation. 
Reducing the block size facilitates a more fine-grained pruning and thus a higher pruning rate.
However, more individual blocks require extra storage for row and column index with the CSB-formatted weight matrix (\reffig{fig:03_csb}).
Therefore, we present both the attainable pruning rate and the index overhead with different block sizes in each benchmark model. 
The block is set to square with sizes of $16$, $32$, $64$, $128$, considering the weight matrix dimensions in different models. 
Note that for matrix with very small size (e.g., $256\times 39$ in \hw{SR4}), the short dimension ($39$) is partitioned to $Q$ blocks uniformly after padding a zero-column.
Multiple layers in one model adopt the same pruning rate.
The attainable pruning rate for each case is presented in \reffig{fig:06_pr1}(a);
Further, the index overheads are divided by the corresponding weight count for normalization, and the values of the \emph{normalized index overhead} (NIO) are presented in \reffig{fig:06_pr1}(b). 
Notably, the results with non-structured pruning are given for comparison (leftmost bar for each application); And its index overhead is obtained by compressing the non-structured weight matrices with the compressed sparse row (CSR) format.

As a result, the CSB pruning rate ranges from $3.5\times$ to $25\times$, which dramatically reduces the original model size by order of magnitude.
With the growth of block size, the pruning rate decreases as the coarse-granularity block reduces the pruning flexibility.
We note that, in all benchmarks, the CSB pruning is capable of reaching a maximum pruning rate with the block size of $16$ or $32$, which is close to non-structured pruning.
In the aspect of NIO, the index overhead of non-structured pruning exceeds $100\%$, as at least one index is required for a non-zero element. 
Nevertheless, for CSB pruning, the NIO is below $50\%$ in most cases due to index reusability in the structured blocks.
The NIO shows a significant decay while enlarging the block size.
With the block size of $32$, the NIO declines to $\approx 20\%$, which is $1/5$ of that in non-structured pruning.
Interestingly, we gain the \embf{insight} that with a block size of $32$ and $16$, most models achieve the close pruning rate. 
For instance, $13\times$ and $12\times$ in \hw{MT1}; both are $20\times$ in \hw{SR2}.
Therefore, the larger block size ($32$) is preferable for its low index overhead.

\subsubsection{Comparison with Prior Art Compression Schemes}
\label{sec:eval_rate_2}

\begin{table}[tb]
\footnotesize
\ra{1.1}
\setlength\tabcolsep{4pt}
\centering
\caption{Pruning Rate Comparison}
\vspace*{-0.07truein}
\begin{tabular}{ll|lllll}
\toprule
\textbf{Abbr.} & \textbf{\begin{tabular}[c]{@{}l@{}}Compression\\ Technique\end{tabular}} & \textbf{\begin{tabular}[c]{@{}l@{}}Prune\\ Rate\end{tabular}} & \textbf{\begin{tabular}[c]{@{}l@{}}Weight\\ Width\end{tabular}} & \textbf{Metric} & \textbf{Result} & \textbf{\begin{tabular}[c]{@{}l@{}}Impro-\\ vement\end{tabular}} \\ \hline
\multirow{2}{*}{\hw{MT1}} & column pruning~\cite{wang2019acceleration} & 8$\times$ & 16-bit & \multirow{2}{*}{PPL} & 112.73 & 1$\times$ \\
 & \textbf{CSB pruning} & \textbf{12.5$\times$} & \textbf{16-bit} &  & \textbf{112.02} & \textbf{1.6$\times$} \\ \hline
\multirow{3}{*}{\hw{MT2}} & row-column~\cite{wen2018learning} & 3$\times$ & floating & \multirow{3}{*}{PPL} & 82.59 & 1$\times$ \\
 & bank balanced~\cite{cao2019efficient} & 5$\times$ & 16-bit &  & 82.59 & 1.65$\times$ \\
 & \textbf{CSB pruning} & \textbf{12$\times$} & \textbf{16-bit} &  & \textbf{82.33} & \textbf{3.9$\times$} \\ \hline
\multirow{4}{*}{\hw{SR1}} & block circulant~\cite{wang2018c} & 8$\times$ & 16-bit & \multirow{4}{*}{PER} & 24.57\% & 1$\times$ \\
 & row balanced~\cite{han2017ese} & 8.9$\times$ & 16-bit &  & 20.70\% & 1.1$\times$ \\
 & bank balanced~\cite{cao2019efficient} & 10$\times$ & 16-bit &  & 23.50\% & 1.3$\times$ \\
 & \textbf{CSB pruning} & \textbf{13$\times$} & \textbf{16-bit} &  & \textbf{20.10\%} & \textbf{1.6$\times$} \\ \hline
\multirow{2}{*}{\hw{SR2}} & block circulant~\cite{li2019rnn} & 8$\times$ & 16-bit & \multirow{2}{*}{PER} & 20.20\% & 1$\times$ \\
 & \textbf{CSB pruning} & \textbf{20$\times$} & \textbf{16-bit} &  & \textbf{20.01\%} & \textbf{2.5$\times$} \\ \hline
\multirow{2}{*}{\hw{SR4}} & column pruning~\cite{gao2018deltarnn} & 14.3$\times$ & 16-bit & \multirow{2}{*}{Accu} & 98.43\% & 1$\times$ \\
 & \textbf{CSB pruning} & \textbf{23$\times$} & \textbf{16-bit} &  & \textbf{99.01\%} & \textbf{1.6$\times$} \\ \hline
\bottomrule
\end{tabular}
\label{tbl:algocomp}
\end{table}
The CSB pruning rate is further compared to the prior art RNN compression techniques in \reftab{tbl:algocomp}.
The listed competitive techniques are proposed to enable a faster, hardware-friendly RNN inference with the compressed model. 
Note that these competitors quantized the weight to $16$-bit fixed-point numbers; 
Thus, we do the same quantization on CSB pruned model and report the corresponding results for a fair comparison.
In \reftab{tbl:algocomp}, \emph{row-column}~\cite{wen2018learning} technique prunes each weight matrix as an entire block. 
Comparing to it, our fine-grained CSB pruning improves the compression rate to $3.9\times$.
The row balanced~\cite{han2017ese} or bank balanced~\cite{cao2019efficient} techniques compulsively train the model to a balanced sparse pattern; 
However, CSB pruning remains the natural unbalanced sparsity in RNN model and achieves a higher ($1.6\times$) pruning rate.
Overall, the CSB pruning improves the pruning rate to $1.6\times$-$3.9\times$ of the existing schemes, while maintaining an even better model accuracy.

\subsection{Evaluation of RNN dataflow Architecture with CSB Pruned Model}
\label{sec:eval_hw}

\subsubsection{Hardware-resource Consumption}

\begin{figure}
  \centering
  \includegraphics[width=0.5\textwidth]{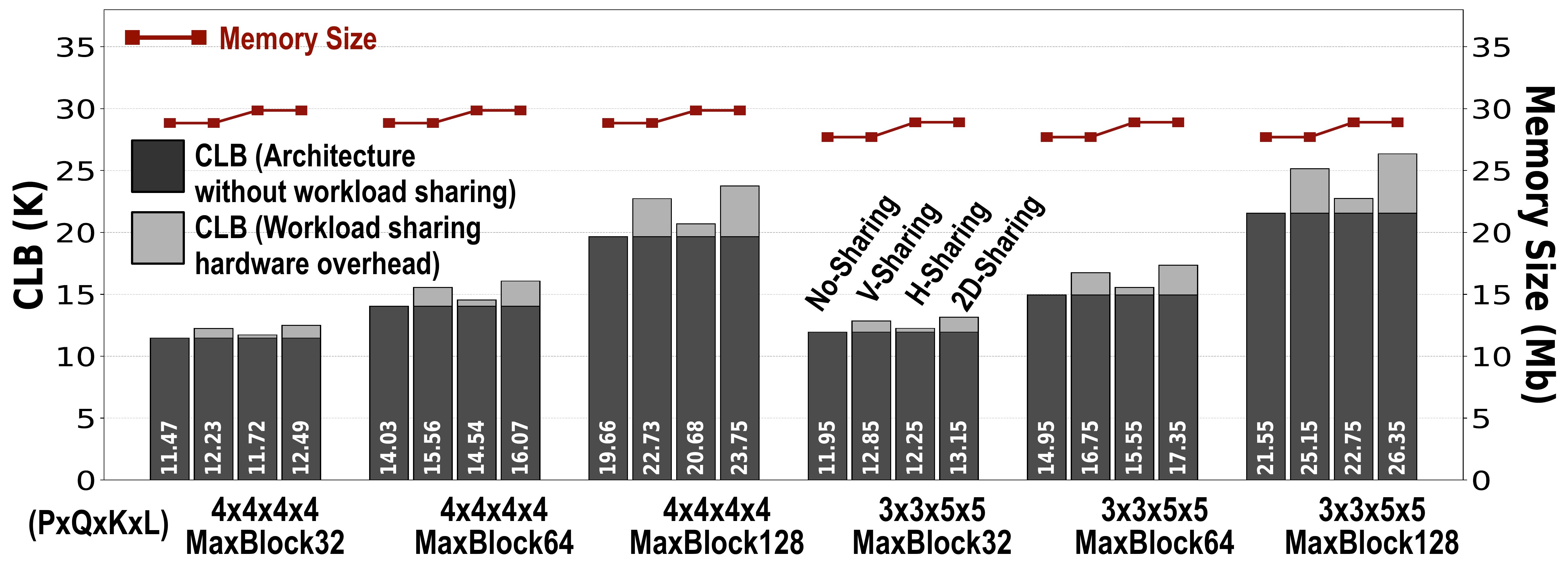}
  \caption{Hardware resource consumption with multi CSB-Engine configs.}
  \label{fig:06_hw}
\end{figure}

The hardware-resource consumption (cost) of the RNN dataflow architecture is given in \reffig{fig:06_hw}, with various \csbengine configs ($P$,$Q$,$K$,$L$ and max supported block size). 
Notably, the \csbengine with different workload sharing configs, including \emph{no-sharing, vertical-sharing, horizontal-sharing, 2D-sharing}, are synthesized individually to evaluate the \embf{hardware overhead} on workload sharing technique.
The consumption of hardware logic and memory from the FPGA vendor tool are presented in \reffig{fig:06_hw}. 
The configurable logic block (CLB, left axis) is the FPGA building block for logic, which is used as the logic resource metric;
The memory resource is given in megabit (Mb in the right axis).
Note that most memory resource on our FPGA device is configured as the weight buffer, although they may not be fully used by small RNN models. 
The multiplier in each \hw{PE} ($16$-bit fixed-point) is mapped to digital signal processor (DSP) on FPGA, and the DSP count in design is $\approx P\times Q\times K\times L$ that is omitted here. 
As a result, the hardware support of workload sharing costs an acceptable overhead, which is $11.6\%$, $3.8\%$, and $15.6\%$ for three sharing cases (vertical/horizontal/2D-sharing), respectively.

\subsubsection{Performance}

\begin{figure*}
  \centering
  \includegraphics[width=1\textwidth]{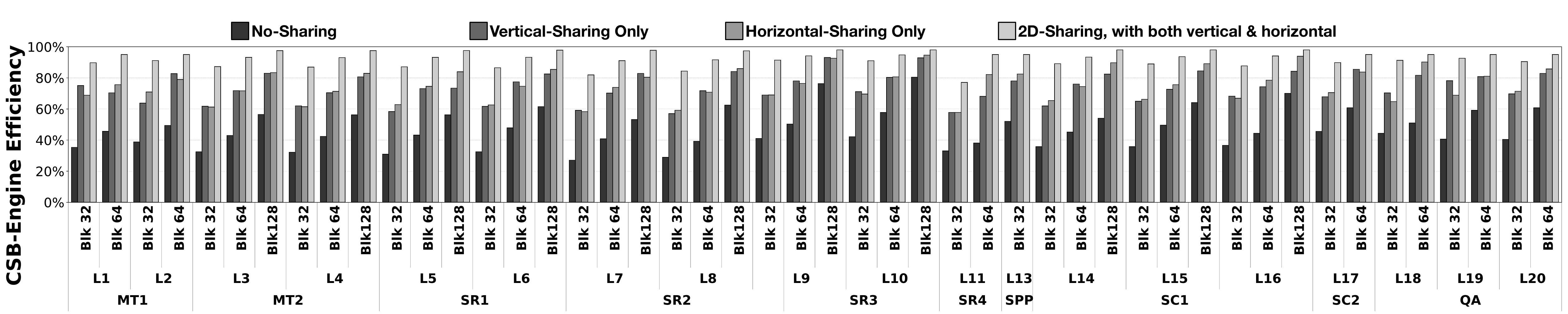}
  \vspace{-2em}
  \caption{The efficiency (utilization) of the proposed architecture with different sharing strategies. The novel workload sharing technique significantly improves the average efficiency from $42\%$ (no-sharing) to $94\%$ (2D-sharing). This improvement fully exploits the benefits of fine-grained CSB pruning.}
  \label{fig:06_eff}
\end{figure*}

Due to the workload imbalance issue, the processing performance of RNN dataflow architecture, \csbengine in specific, is not deterministic. 
Hardware efficiency, the ratio of effective computation on \hw{PE}s, is invoked to evaluate the improvement of our workload sharing technique. 
We obtained the \csbengine efficiency by measuring the \hw{PE} pipeline utilization using $10$ benchmarks listed in \reftab{tbl:benchmark} with different design choices of workload sharing.
Moreover, CSB pruned models with different block sizes are used to evaluate the impact of block size on efficiency.
The efficiency is measured layer-by-layer on hardware with $4\times 4$ \hw{PEGroup}s and each contains $4\times 4$ \hw{PE}s. 
The results are presented in \reffig{fig:06_eff}.
Overall, for the \csbengine without workload sharing, the efficiency is $42\%$ on average, which results from the imbalanced workload (sparsity) of blocks.
The single dimensional sharing (vertical or horizontal) improves the efficiency to an average of $72\%$. 
After the 2D-sharing is adopted, the efficiency is further improved to $94\%$ on average, i.e., only $6\%$ execution time of \csbengine is invalid.
This $6\%$ pipeline gap is inevitable, as a few extremely imbalanced sparsity exists in some weight matrices.
For instance, we found \embf{diagonal dense matrix} exists that the blocks on the matrix diagonal contain significant workload compared to other blocks.
In this case, the workload sharing path in the current design is not enough, while adding more sharing paths brings extra hardware costs. 

Comparing the efficiency within the same layer but different pruning block sizes, it is apparent that the smaller block size is applied, the lower hardware efficiency \csbengine can achieve, particularly in the no-sharing \csbengine cases.
This is because the small block includes less workload (with the same pruning rate) but more temporal block iterations, which lead to \hw{PE} idle more easily. 
As mentioned in \refsec{sec:eval_rate_1}, using smaller block sizes in compression guarantees higher model pruning rates, which benefits are significantly encroached by the performance degradation with small compression block in the no-sharing cases.
Nevertheless, we gain the \embf{insight} that our architecture-compilation co-design for 2D-sharing cases significantly subdues the degradation.
For instance, in Layer-2 (L2) of \hw{MT1} case, the no-sharing degradation from block-64 to block-32 is $12\%$, while it is reduced to $3\%$ by the 2D-sharing. 
On average, the degradation is reduced from $15\%$ to $4\%$.
In summary, with the proposed workload sharing technique, a smaller block size in CSB pruning does not bring significant degradation on hardware efficiency anymore (only $4\%$ on average), so that the benefits from higher pruning rates can be more sufficiently exploited.

\subsubsection{Comparison with Related Works}

The overall performance of CSB-RNN, i.e., CSB pruned model inference speed on the proposed RNN dataflow architecture, is listed in \reftab{tbl:hwcomp} and compared with the prior art designs. 
We collected the statistics including the PE count (\#PE), operating frequency, latency in processing one input frame and the power of design. 
As \reftab{tbl:hwcomp} shows, with the same benchmark applications, the CSB-RNN reduces the latency by $39\%$-$92\%$ that speeds up the processing from $1.12\times$ to $12.57\times$ correspondingly;
Nevertheless, CSB-RNN only uses $19\%$-$34\%$ PE counts (hardware resource) of the competitors to attain this performance.
The latency ranges from $0.79\mu$s to $6.58\mu$s with different model sizes. 
For generic high-precision speech recognition, at most $\approx 2000$ frames should be processed per second, which requires a latency $\leq 500\mu$s to meet the \emph{realtime} performance.
As the achieved latency with benchmark models is much lower than this requirement, the CSB-RNN provides a \embf{faster-than-realtime} performance and facilitates the device processing more complex RNN models in the future.
Besides the latency, we compare the \embf{power efficiency} (k-frames per Watt) among these competitive designs.
The results show the CSB-RNN achieves significant improvements from $3.53\times$ to $58.89\times$ on power efficiency in processing the same model, which makes the CSB-RNN quite suitable for embedded scenarios.
Further, while the existing works were designed for a particular RNN cell type, CSB-RNN can be reprogrammed to adapt to different cells.

\begin{table}[tb]
\footnotesize
\ra{1.1}
\setlength\tabcolsep{2.1pt}
\centering
\caption{Latency and Power Efficiency Comparison}
\vspace{-1em}
\begin{tabular}{cc|cc|ll|ll}
\toprule
\multirow{2}{*}{\textbf{Abbr.}} & \multirow{2}{*}{\textbf{Work}} & \textbf{\#PE} & \textbf{Freq.} & \multicolumn{1}{c}{\textbf{Latency}} & \multicolumn{1}{c|}{\textbf{Power}} & \multicolumn{1}{c}{\textbf{Power Eff.}} & \multicolumn{1}{c}{\textbf{Power Eff.}} \\
 &  & \textbf{} & (MHz) & \multicolumn{1}{c}{($\mu$s)} & \multicolumn{1}{c|}{(Watt)} & \multicolumn{1}{c}{(k-frames/W)} & \multicolumn{1}{c}{\textbf{Improv.}} \\ \hline
\multirow{2}{*}{$\mathtt{MT1}$} & BBS~\cite{cao2019efficient} & 1518 & 200 & 1.30 & 19 & 40.49 & 1$\times$ \\
 & \textbf{CSB-RNN} & \textbf{512} & \textbf{200} & \textbf{0.79} & \textbf{8.9} & \textbf{142.72} & \textbf{3.53$\times$} \\ \hline
\multirow{4}{*}{$\mathtt{SR1}$} & C-LSTM~\cite{wang2018c} & 2680 & 200 & 8.10 & 22 & 5.61 & 19.35$\times$ \\
 & E-RNN~\cite{li2019rnn} & 2660 & 200 & 7.40 & 24 & 5.63 & 19.41$\times$ \\
 & ESE~\cite{han2017ese} & 1504 & 200 & 82.70 & 41 & 0.29 & 1$\times$ \\
 & \textbf{CSB-RNN} & \textbf{512} & \textbf{200} & \textbf{6.58} & \textbf{8.9} & \textbf{17.08} & \textbf{58.89$\times$} \\ \hline
\multirow{2}{*}{$\mathtt{SR2}$} & E-RNN~\cite{li2019rnn} & 2280 & 200 & 6.70 & 29 & 5.15 & 1$\times$ \\
 & \textbf{CSB-RNN} & \textbf{512} & \textbf{200} & \textbf{5.18} & \textbf{8.9} & \textbf{21.69} & \textbf{4.21$\times$} \\ \hline
\bottomrule
\end{tabular}
\label{tbl:hwcomp}
\end{table}

\section{Conclusion}
This paper presents CSB-RNN, an optimized full-stack RNN acceleration framework. 
The fine-grained structured CSB pruning significantly improves the pruning rate compared to existing hardware-friendly pruning schemes.
Meanwhile, an architecture-compilation co-design is proposed that sufficiently exploits the benefits of the CSB pruned model. 
The experiments show that the entire CSB-RNN acceleration framework delivers a faster-than-realtime performance on extensive RNN models, and dramatically reduces the latency and improves the power efficiency compared with the existing works. \\
\textbf{Future work:} We are extending the CSB technique to other neural network layers. In particular, the transformer models are composed of more complex dataflow, however, the same MVM primitive as RNN. With improvement on the dataflow abstraction, the proposed CSB pruning and \csbengine  will contribute to the realtime transformer inference. 

\begin{acks}
We would like to thank the anonymous reviewers for their valuable comments.
This research was supported in part by the Croucher Foundation (Croucher Innovation Award 2013), the Research Grants Council of Hong Kong grant number CRF C7047-16G, GRF 17245716.
This research was supported in part by the U.S. DOE Office of Science, Office of Advanced Scientific Computing Research, under award 66150: ``CENATE - Center for Advanced Architecture Evaluation''.
This research was supported, in part, by the NSF through awards CCF-1618303, CCF-1919130, CCF-1937500, CNS-1909172, and CCF-1919117; the NIH through awards 1R41GM128533 and R44GM128533; and by a grant from Red Hat.

\end{acks}

\bibliographystyle{ACM-Reference-Format}
\bibliography{ref}

\end{document}